\documentstyle[12pt,amsfonts,epsf]{article}

\begin{document}

\def\R{\mathbb R}
\def\be#1{\begin{equation}  \label{#1}}          
\def\ee{\end{equation}}                          
\def\nor#1{\vert #1 \vert}  
\def\intdrr{\int\!\!\!\!\int\limits_{\!\!\!\Omega}\!\!}

\title{Morphological Model for Colloidal Suspensions} 

\author{Uwe Brodatzki and Klaus R. Mecke \\
Fachbereich Physik, Bergische Universit\"at Wuppertal \\  
 D - 42097 Wuppertal, Federal Republic of Germany  }

\maketitle

\bigskip 
\bigskip 
\bigskip 

\noindent 
Paper presented at the 14th Symposium on Thermophysical Properties,
June 25-30, 2000, Boulder, Colorado, USA. 

\bigskip 
\bigskip

\newpage

\begin{abstract}
The phase behavior of colloidal particles embedded in a binary fluid
is influenced by  wetting layers surrounding each particle. The free energy of
the  fluid film depends  on its morphology, i.e.,  on   size, shape and 
connectivity.   
Under rather natural assumptions a general expression for the
Hamiltonian can be given extending the
model of hard spheres to partially penetrable shapes including energy
contributions related to the volume, 
surface area, mean curvature, and Euler characteristic of the wetting
layer. 
The complex spatial structure  leads to multi-particle  
  interactions of the colloidal particles. 

The dependence of the morphology of the wetting layer on temperature
and density  can be studied using Monte-Carlo
simulations and  perturbation theory. A fluid-fluid phase
separation induced by the  wetting layer is observed  which is  
suppressed when only two-particle interactions are taken into
account instead of the inherent 
many-particle interaction  of the wetting layer. 

\end{abstract}

\bigskip 
\bigskip 
\bigskip


\newpage 

\section{Introduction}
\label{introduction}

Colloidal solutions such as paints and
soots are extremely common and  exhibit  many industrial
and technological applications.  Due to 
their mesoscopic size they are also of great fundamental 
importance  and have provided basic parts of our understanding
of the interaction of particles, for instance.    
In most cases the electrostatic repulsive interaction is significantly
screened so that the attractive dispersion interaction 
dominate which finally cause an irreversible aggregation or coagulation. The
colloidal particles stick, i.e., are in point contact at the global minimum of the
interaction potential. But under certain circumstance  reversible
colloidal aggregation or flocculation can be 
observed.   Systematic experiments 
\cite{beysens,gallagher,jayalakshmi} on colloidal
particles embedded in a near-critical solvent mixture of 2,6-lutidine
plus water, for instance, have revealed that flocculation can be viewed as thermally
induced phase separation. 

Theoretical work of colloidal partitioning or flocculation in a
two-phase solvent focused on general thermodynamic arguments known
from wetting or capillary condensation \cite{beysens2,dobbs}
and more recently on a Ginzburg-Landau approach including a detailed
treatment of the first-order solvent phase transition which is coupled
to the colloidal particles \cite{loewen,gil}. 
These works are based  on microscopic pair interactions
between the colloidal particles and the  molecules in the 
binary fluid. Thus, the colloids do not interact
directly, but couple 
to the binary fluid  degrees of freedom by preferring one of the two
components. 

The phase behavior of colloidal particles embedded in a binary fluid
is  certainly 
influenced by the existence of  wetting layers, i.e. by  thermodynamically
meta-stable fluid phases  stabilized at the boundary of the colloidal
particles. Wetting phenomena  appear in
multicomponent systems when the 
components exhibit different interactions with the colloidal
particle. 
In general, when two thermodynamic phases are close to a first order
phase transition  the wetting 
layer may become large and comparable to the diameter of the colloidal
particle.  
The interactions  are then determined 
by  the free energy of the  fluid film between   the hard colloidal
particles which cause clustering and eventually  a 
phase separation.

Here, we focus on an effective theory where the binary fluid enters
only in one parameter, namely the thickness $L(T,c)$ of the wetting
layer (enriched layer of one component) which depends on temperature $T$ and
concentration $c$ of one component of the binary fluid. 
In other words, the colloidal system and the binary fluid are coupled 
solely by  the parameter  $L(T,c)$ which can be determined for a
single colloidal particle in a 
binary fluid without taking into account the interaction with other
particles.  
The advantage of this approach is the computational simplicity of the
influence of the 
binary fluid on the interaction of the colloidal particles. This
simplicity 
allows a more sophisticated treatment of the colloidal interaction
itself. Taking into account not only pair-potentials one can study,
for instance, 
the effect of multiple-particle interactions due to an overlap of the
wetting layer of many particles. This cannot be neglected when $L$ is
large in the complete wetting regime, i.e., close to the critical
point of the binary fluid. 

Since 
the thermodynamic properties of the fluid film depend  on its 
  morphology, i.e., on  volume and surface
area, a statistical theory should include geometrical descriptors
to characterize the size, shape and 
connectivity of the wetting layer. 
In Section \ref{measures} we proposed a model for the study of such colloidal
suspensions.  
The colloids are resembled by spherical particles (disks in two
dimensions) with a hard-core 
diameter $D$ and a soft, penetrable shell of thickness $L$ (see Figure
\ref{decorated}). Using the radius $R=D/2+L$ of the partially penetrable
disks  one can define the ratio $\delta=D/(2R)$ with  $0 \leq \delta
\leq 1$, where $\delta=1$ equals a pure hard-disk system, whereas
$\delta=0$ denotes an ensemble of fully penetrable disks.    The
Hamiltonian includes not only the  two-particle hard-sphere 
 potential but also energy contributions related to the volume,
surface area, mean curvature, and Euler characteristic of the 
layers around the hard particles.    

In Section \ref{perturbation} a perturbation theory and in Section \ref{montecarlo} a Metropolis Monte-Carlo algorithm  of partially penetrable
disks in two dimensions is  
presented. 
The morphology of the wetting layer as function of temperature and
density of the colloidal particles as well as 
phase diagrams are discussed.

\section{Morphological thermodynamics of colloidal configurations}
\label{measures}

From a geometric point of view one should tie the interaction of the
 colloidal particles to the  morphology of the wetting layer, for instance,
to its  volume and surface area.  
Each configuration 
\be{conf}
{\cal A}^N=\bigcup_{i=1}^NB_R^D(\vec{x}_i) \;\;\;.
\ee
is assumed to be the
union of mutually penetrable $d$-dimensional spheres $B_R^D(\vec{x})$ of
 radius $R$ and hard-core diameter $D$ 
 centered at $\vec{x}\in \Omega \subset {\R}^d$  
embedded in binary fluid 
host component. For convenience, we assume a box $\Omega$ of edge
 length $1$ and periodic boundary conditions so that all lengths are
 measured in units of the box 
 size.  
Typical configurations are shown in Figure \ref{decorated} for
 $\delta=D/(2R)= 0.4$. In the following we consider two
 dimensional systems with penetrable disks. 
Depending on the density $\rho$ of the particles and the size $L$ of
 the soft shell, i.e. the ratio 
$\delta=1/(1+2L/D)$    
of the radii, the wetting layer (i.e., the white area in Figure
 \ref{decorated}) exhibit quit different topological and 
geometric properties. For instance, the white disk shells are  disconnected
(isolated) for $\delta \rightarrow 1$  due to the hard
core interaction, whereas for 
small $\delta \approx 0$ the grains can    
overlap so that at higher densities $\rho$   connected structures occur. 
The morphology of the emerging pattern may be characterized by the
covered volume or the area in two dimensions, the surface area or boundary
length, respectively,  and the Euler-characteristic, i.e., the connectivity of the
penetrating grains.   
Area, boundary length and  Euler characteristic have in
common that they are additive measures.  
Additivity means that the measure of the union $A\cup B$
of two  domains (grains) $A$ and  $B$ equals the sum of the
measure of the single domains subtracted by the  
intersection, i.e., 
\be{additivity}
{\cal M}(A\cup B)={\cal M}(A)+
{\cal M}(B)-{\cal M}(A\cap B)\; .  
\ee 
A remarkable theorem in integral geometry
\cite{santalo,hadwiger} is the  completeness  of
the so-called Minkowski functionals. The theorem  asserts that any 
additive, motion invariant and conditional continuous
functional ${\cal M}(A) = \sum_{\nu=0}^{d} c_{\nu}M_{\nu}(A)$ on
 subsets $A \subset \R$$^d$,  is a linear  
combination of 
the $d+1$ Minkowski functionals $M_\nu$, ($\nu=0,\ldots,d$)  
with real coefficients $c_\nu$ independent of $A$. The two conditions
of motion invariance and conditional continuity are necessary for the
 theorem, but they are not  
very restrictive in most physical situations. Intuitively,  conditional 
continuity  expresses the fact that 
an approximation of a convex domain $K$ by convex polyhedra $K_n$, for
example, also yields an approximation of ${\cal M}(K)$ by
${\cal M}(K_n)$. The property of
motion-invariance means that  
the morphological measure $\cal M$ of a domain $A$ is independent of its
location and orientation in space. 
Thus, every
morphological measure $\cal M$ which is additive (motion-invariant
and continuous), i.e., which obeys 
relation (\ref{additivity})   
can be written in terms of  
Minkowski functionals $M_\nu$, which are related to curvature 
integrals  and do  not only
characterize the size   but also shape (morphology) and connectivity (topology)
 of spatial patterns.  
In the 
three-dimensional Euclidean space the family of Minkowski functionals
consists of  the volume $V=M_0$, the  
surface area $S=8M_1$ of the 
pattern,  its integral mean curvature $H=2\pi^2M_2$, and integral of
Gaussian curvature. In two dimension
 they are given by the area $F=M_0$, the  length $U=2\pi M_1$ and the
 curvature integral, i.e., 
the   Euler characteristic $\chi = \pi M_2$  of the boundary.   
In other words,  the Minkowski functionals are the 
complete set of additive, morphological measures. We assume, that the 
energy of a configuration is  a
morphological measure of the wetting layer, i.e., an additive
 functional of the fluid films surrounding each particle.   
Thus, the Boltzmann weights are specified by the Hamiltonian 
\be{thermo1}
{\cal H}({\cal A}^N) = \sum_{i\neq j =1}^N V(\vec{x}_i,\vec{x}_j) \;  +
\; \sum_{\nu=0}^d h_\nu \left(M_\nu\left(\bigcup_{i=1}^N
B_R^D(\vec{x}_i)\right) - NM_\nu(B_R^D) \right) 
\ee
which is a linear combination of Minkowski functionals of the
configurations and a pair potential 
\be{pairpot}
V(\vec{x}_i,\vec{x}_j) = \left\{\matrix{ 0 & |\vec{x}_i-\vec{x}_j| > D
\cr 
\infty & |\vec{x}_i-\vec{x}_j| \leq D
\cr }\right. 
\ee
between two colloidal particles located at $\vec{x}_i$ and
$\vec{x}_j$. For convenience we assume a pure 
hard-core interaction.  
We emphasize that the Hamiltonian
(Eq. (\ref{thermo1}))  
constitutes  the 
most general model for composite media assuming 
additivity of the free energy of the homogeneous wetting layer. 
 The interactions between the colloidal particles are given by a
bulk term (volume energy), a surface term (surface tension), and
curvature terms (bending energies) of the wetting layer.

We define the packing fraction
$\eta={\pi\over 4} D^2 \rho  = x\delta^2$ 
and the normalized density $x=\rho \pi R^2$ 
of the disks.  
The closest packing fraction is $\eta_{CP}= {\pi \over 2\sqrt{3}}$ or
$x_{CP}=\eta_{CP}/\delta^2$, i.e.,  $x_{CP}(\delta=0.4)\approx
5.67$. Since we are not interested in the solid-liquid phase
transition of hard disks but in a fluid-fluid transition induced by the
wetting layer we focus on densities well below $x_{CP}$.  
Depending on the ratio $\delta$ it is possible to have
multiple-overlapps of disks. In particular, two disks can overlapp 
iff $\delta <1$, three iff $\delta < \sqrt{3}/2\sim 0.866$, and four iff
$\delta < 1/\sqrt{3}\sim 0.577$.  If $\delta < 1/2$ a proliferation of
possible overlapps occur.   
In this limit a virial expansion of the  Minkowski functionals  in
the 
density is not useful anymore and the interaction has manifestly
many-body character.   Therefore, we focus in this paper on $\delta=0.4$ in
comparision with $\delta=1$ and $\delta=0$ as limiting cases.

\section{Perturbation theory} 
\label{perturbation} 

Because of the proliferation of multibody potentials, an exact
evaluation of the partition function for the Hamiltonian
(\ref{thermo1})  appears to be 
unmanagable for $d \geq 2$.  
For the sake of comparison with Monte-Carlo simulations presented in
the next section, we calculate here the thermodynamic properties and
phase diagram of the model by simple first-order thermodynamic
perturbation theory \cite{hansen}. This approximation keeps 
the geometrical and topological aspects of the model intact. As a
reference system, we use the 
hard-sphere fluid and solid. The free energy is estimated as 
\be{freenergy}
\beta F(T,\rho) = \beta F_{HS}(\rho) + \beta <U>_{HS}(\rho)
\ee
where $F_{HS}(\rho)$ is the free energy of the hard-sphere system at
density $\rho$, and $<U>_{HS}(\rho)$ is the average value of the
morphological part of the Hamiltonian (\ref{thermo1}), computed in the
hard-sphere reference system. Thus, keeping only the first two
terms in a high-temperature expansion of the free energy  amounts to
replacing the configurational integral  in the 
partition function  by
$\exp\left\{-\beta <{\cal U}>_{HS}\right\}$ 
which yields a lower bound.  
The equation of state for hard disks is given by  the  Pad\'{e}
approximation from Hoover and Ree (1969) for the free 
energy \cite{hoover} 
\be{pade1}
\frac{\beta F_{HS}}{N}
= \log \left(\rho\Lambda^2\right)  -1 + b\rho \frac{1-0.28 b\rho +
0.006b^2\rho^2}{1-0.67 b\rho + 0.09 b^2\rho^2}
\ee
where $\Lambda$ is the mean thermal de Broglie
wavelength and  $b=\frac{\pi}{2} D^2$. 
The packing fraction  above the freezing
transition, i.e., for large values of $\rho$,
can be calculated by the  free volume theory.

The perturbation energy 
$<U>_{HS}(\rho) =  
\sum_{\nu=0}^d h_\nu \left(\bar{M}_\nu(x)  - 
NM_\nu(B_R^D) \right)$ in Eq. (\ref{freenergy}) is given by the
average values  $\bar{M}_\nu(x)$ 
of the Minkowski functional in the hard-sphere reference system. 
In general, the average values of the Minkowski functionals
for correlated disks of radius $R$ 
are given  by 
\be{minkgeneral}
\begin{array}{rl} 
\bar{M}_0(x) = & 1- e^{-x f(x,c^{(n)}) }\;,   \cr
\bar{M}_1(x) = & x  u(x,c^{(n)})e^{-x f(x,c^{(n)}) } \;,  \cr 
\bar{M}_2(x) = & x \left(e(x,c^{(n)}) - x \left(u(x,c^{(n)})\right)^2
 \right)e^{-x f(x,c^{(n)}) } \;,   \cr   
\end{array} 
\ee
where the functions $f(x,c^{(n)})$, $u(x,c^{(n)})$ and $e(x,c^{(n)})$
depend on the normalized density of disks 
$x=\rho \pi R^2$ and the hierarchy of correlation functions 
$c^{(n)}(\vec{x}_1,\ldots,\vec{x}_n)$ of the centers of the disks. 
In lowest order, i.e., for an approximation by a Gau\ss -Poisson process one
obtains the correlated average of the 
intersectional area, boundary length, and integral curvature in terms
of the two-point correlation function $g(\vec{r})$ with
$c^{(2)}(\vec{r}) = (g(r)-1) \rho^2$: 
\be{minkgeneral2}
\begin{array}{rl} 
f(x) =  & 1- {1 \over 2\rho} \int\limits_{\R^d} d\vec{r}
c^{(2)}(\vec{r}) \left(1 - {2 \over \pi} \arcsin {|\vec{r}| \over 2R}  - {2
\over \pi}{|\vec{r}| \over 2R} \sqrt{1-\left({|\vec{r}| \over 2R}\right)^2}\right) \cr 
u(x) =  & 1- {1 \over 2\rho}  \int\limits_{\R^d} d\vec{r}
c^{(2)}(\vec{r}) \left(1 - {2 \over \pi} \arcsin {|\vec{r}| \over 2R}  \right)  \cr 
e(x) =  &1  - {1 \over 2\rho} \int\limits_{\R^d} d\vec{r}
c^{(2)}(\vec{r})\Theta(2R-|\vec{r}|) \;\;.  \cr 
\end{array} 
\ee  
For a Poisson distribution with  $g(\vec{r})=1$ one obtains
$f^{(P)}(x,c^{(n)}) = u^{(P)}(x,c^{(n)}) = e^{(P)}(x,c^{(n)}) = 1$.  
For a hard-core process approximated by the correlation function the
mean values 
\be{hardcore}
g(\vec{r}) = \left\{\matrix{0  & |\vec{r}| < D \cr 
1 & |\vec{r}| \geq D \cr}\right. 
\ee
one obtains  with $\delta=D/(2R)<1$  
\be{minkhardcore1}
\begin{array}{rl} 
f^{(hc)}(R,D) =  & 1+ {x \over 2}\left(1- {2 \over \pi}
\delta\sqrt{1-\delta^2}(1+2\delta^2)  -{2 \over
\pi}(1-4\delta^2)\arccos \delta \right)  \cr
  & \cr  
u^{(hc)}(R,D) =  & 1+ x \left(1- {2 \over \pi}
\delta\sqrt{1-\delta^2}  - {2 \over \pi}(1-2\delta^2)\arccos\delta
\right)   \cr 
  & \cr
e^{(hc)}(R,D) =  & 1+ 2x \delta^2  \;.   \cr  
  & \cr  
\end{array} 
\ee  
From Eqs. (\ref{freenergy}), (\ref{pade1}), (\ref{minkgeneral}), and
(\ref{minkhardcore1}) it is possible to 
derive all thermodynamic properties of the system needed to construct
the phase diagram. 
In the limit $\delta=0$ one recovers the 
 equation of state   for an ideal
gas of overlapping 
disks.

\section{Monte-Carlo algorithm for overlapping disks} 
\label{montecarlo} 

\begin{figure}
\centerline{\Huge Figure 1}
\vskip 4cm
\hbox{\setlength{\epsfxsize}{7cm}
\epsfbox{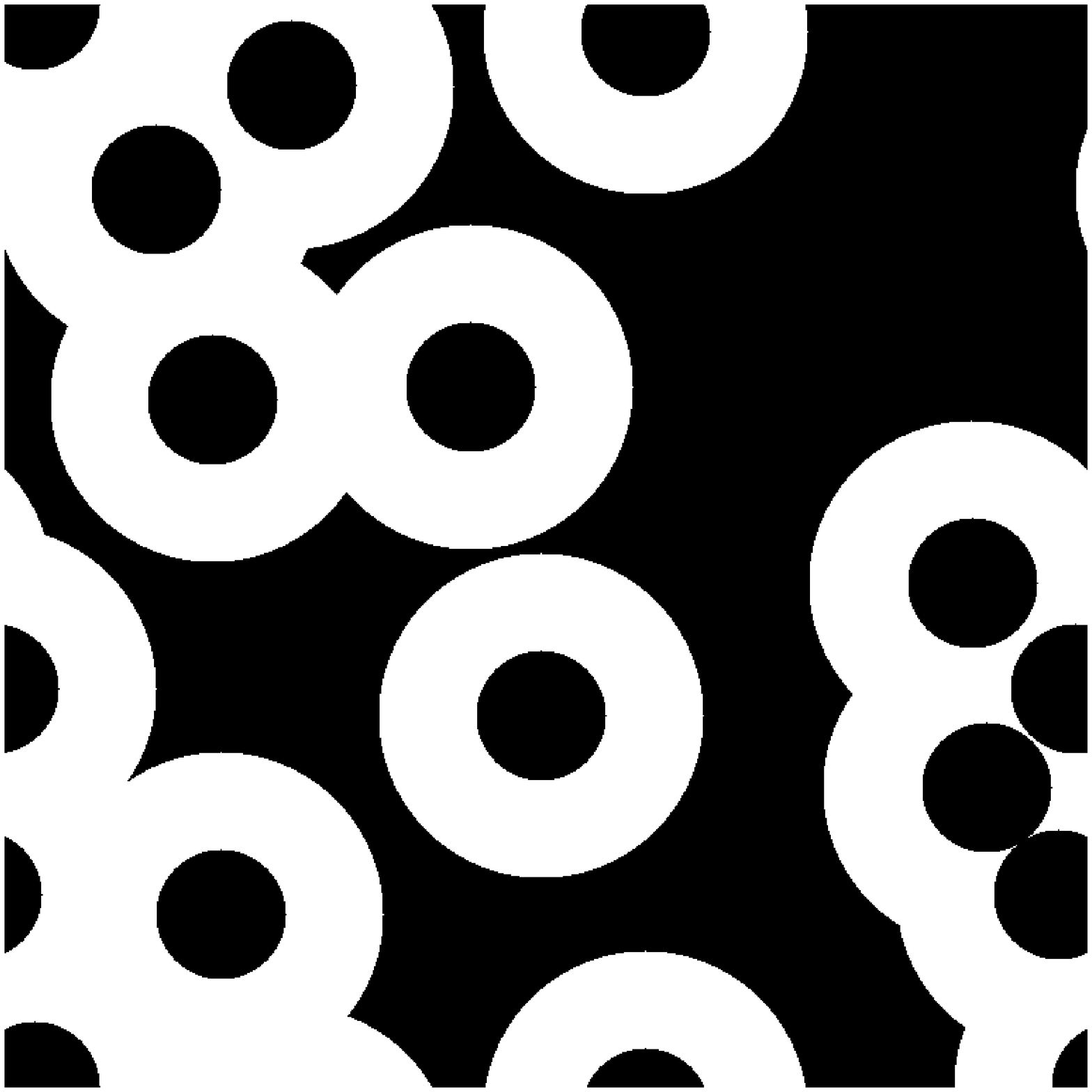}   
\hskip 1cm  
\setlength{\epsfxsize}{7cm}
\epsfbox{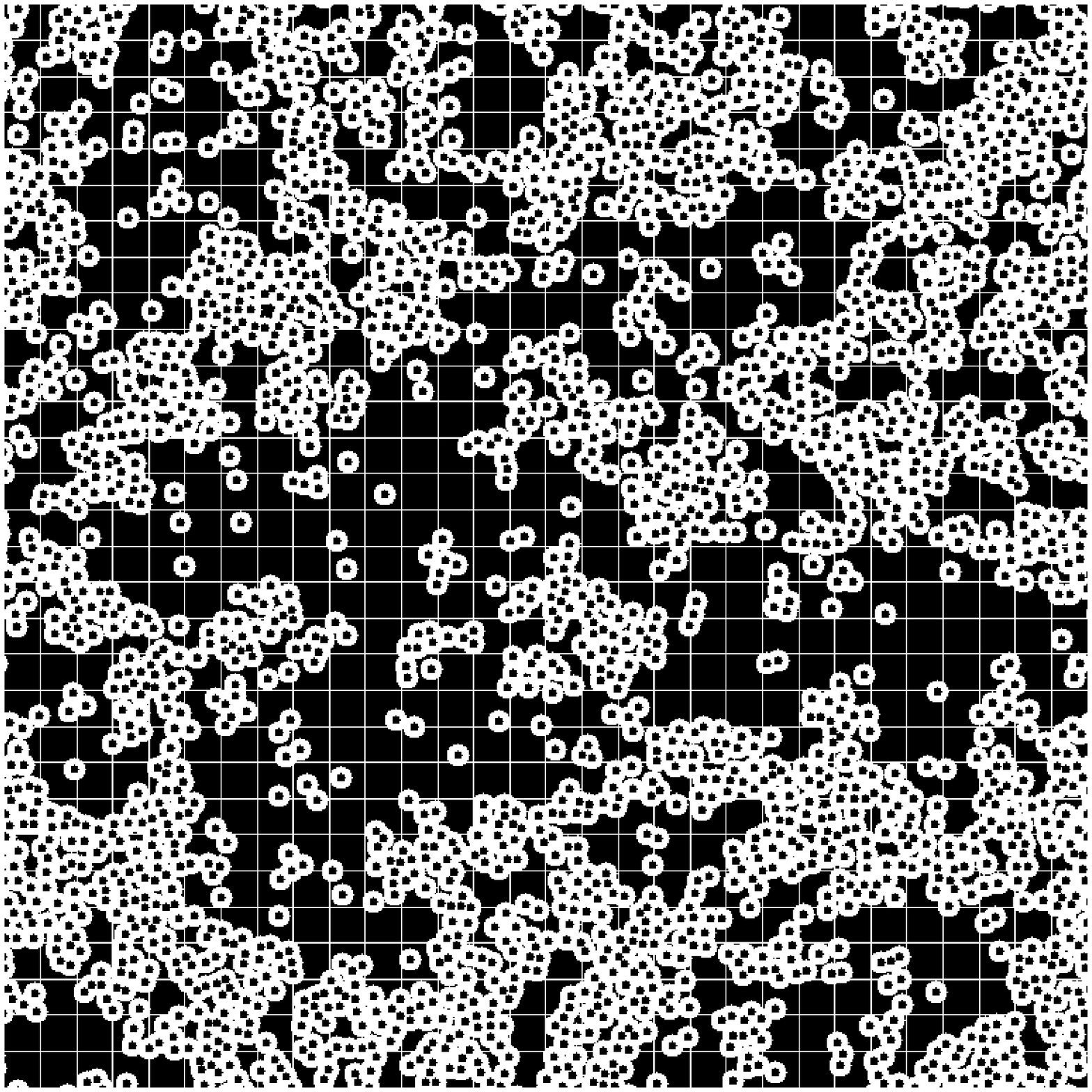} }
\vskip 2cm
\caption{
U. Brodatzki, K. Mecke, Morphological Model for Colloidal Suspensions}  
\label{decorated} 
\end{figure}

With the definition (\ref{thermo1}) of the Hamiltonian ${\cal H}({\cal A})$
for a configuration ${\cal A}=\cup_iB_R^D(\vec{x}_i)$  of  partially
penetrable disks $B_R^D(\vec{x}_i)$ it is 
possible to perform  directly 
Monte-Carlo simulations. Although  lattice
models based on the morphological Hamiltonian (\ref{thermo1}) were
defined and extensive   
computer simulations have been performed  for complex fluids such as
colloidal dispersions and  microemulsions \cite{likos},  not many continuum 
models are studied yet. 
A model based on  correlated partially overlapping grains (see Figure
\ref{decorated}) seems to be a promising starting point to study 
geometric features of complex fluids. In this
section we describe the implementation of an algorithm in two
dimensions for the parallel machine CM5 using the SIMD (single
instruction multiple data) technique. 

In a first step,  
the Minkowski measures $M_{\nu}({\cal A})$  are calculate analytically
for a given configuration 
${\cal A}^N=\cup_{i=1}^{N} B_R^D(\vec{x}_i)$ of $N$ partially
overlapping disks. Then, discs are added, removed, or shifted which
leads to 
a new configuration 
${\cal A}'$. The difference in the Minkowski measures, i.e., in the
Hamiltonian ${\cal H}$  determines the
probability
for the acceptance of the new configuration ${\cal A}'$ according to
the usual Metropolis dynamics.  
Unfortunately, the computational cost to evaluate the energy (\ref{thermo1})
of a configuration  given by Eq. (\ref{thermo1}) is enormous so that
an efficient algorithms is necessary in order to make Monte-Carlo
simulations feasible.

The Minkowski functionals $M_{\nu}({\cal A}^N)$ for the union
$\cup_{i=1}^NB_R^D(\vec{x}_i)$ 
  may be calculated
straightforwardly via the  additivity relation (\ref{additivity}),
i.e., as a sum 
of multiple overlaps  
\be{addi2}
M_{\nu} ({\cal A}^N ) = \sum_i M_{\nu} (B_i)
- \sum_{i<j} M_{\nu} (B_i \cap B_j)  
 +  \ldots + (-1)^{N+1}
M_{\nu} (B_1 \cap \ldots \cap B_N )\;\;. 
\ee
which follows from Eq. (\ref{additivity}) by induction.
The right hand side of Eq. (\ref{addi2}) only involves convex
sets and may be applied together with the definition of $M_{\nu}$
to compute $M_{\nu}({\cal A}^N)$.  
 However, this algorithm becomes inefficient when the amount
of overlap between the augmented balls is excessive, since one has to
compute many redundant and mutually cancelling terms.
Thus, this approach works only for $\delta \approx 1$. 
Therefore, we proceed alternatively as described
in Ref. \cite{meckebuch}. 
The morphology of a configuration $\cal A$ (see Figure \ref{decorated}) is 
unambiguous determined through the borderline between black and white
regions. 
So it is possible to determine all Minkowski measures $M_\nu$
with an appropriate parameterization together
with the local curvature of those borderlines. 
For instance, the two-dimensional area $M_0$ can be calculated by
applying   Gauss' theorem 
\be{vol_gauss}
M_0 ({\cal A})= {1 \over 2}\int\limits_{\partial {\cal A}} {\bf x}\cdot{\bf n}\,dS
\ee   
where $\vec{n}$ denotes the normal vector to the boundary $\partial
{\cal A}$ at $\vec{x}$.  
Accordingly, one obtains the length of the boundary line  $M_1 ({\cal
A})=\int\limits_{\partial {\cal A}} dS$ and the Euler characteristic $M_2 ({\cal
A})= M_1 ({\cal A})/(\pi R) + \sum_i \chi_i/(2\pi)$ 
as integrals along the boundary $\partial {\cal A}$ of a wetting layer
shown in Figure
\ref{decorated}, for instance. Here, $\chi_i$ denotes the angle between
the two normals  at  
the intersection point $i$ 
of two disk boundaries.

\begin{figure}
\centerline{\Huge Figure 2}
\vskip 4cm
\hbox{\epsfxsize 7.cm
\epsffile{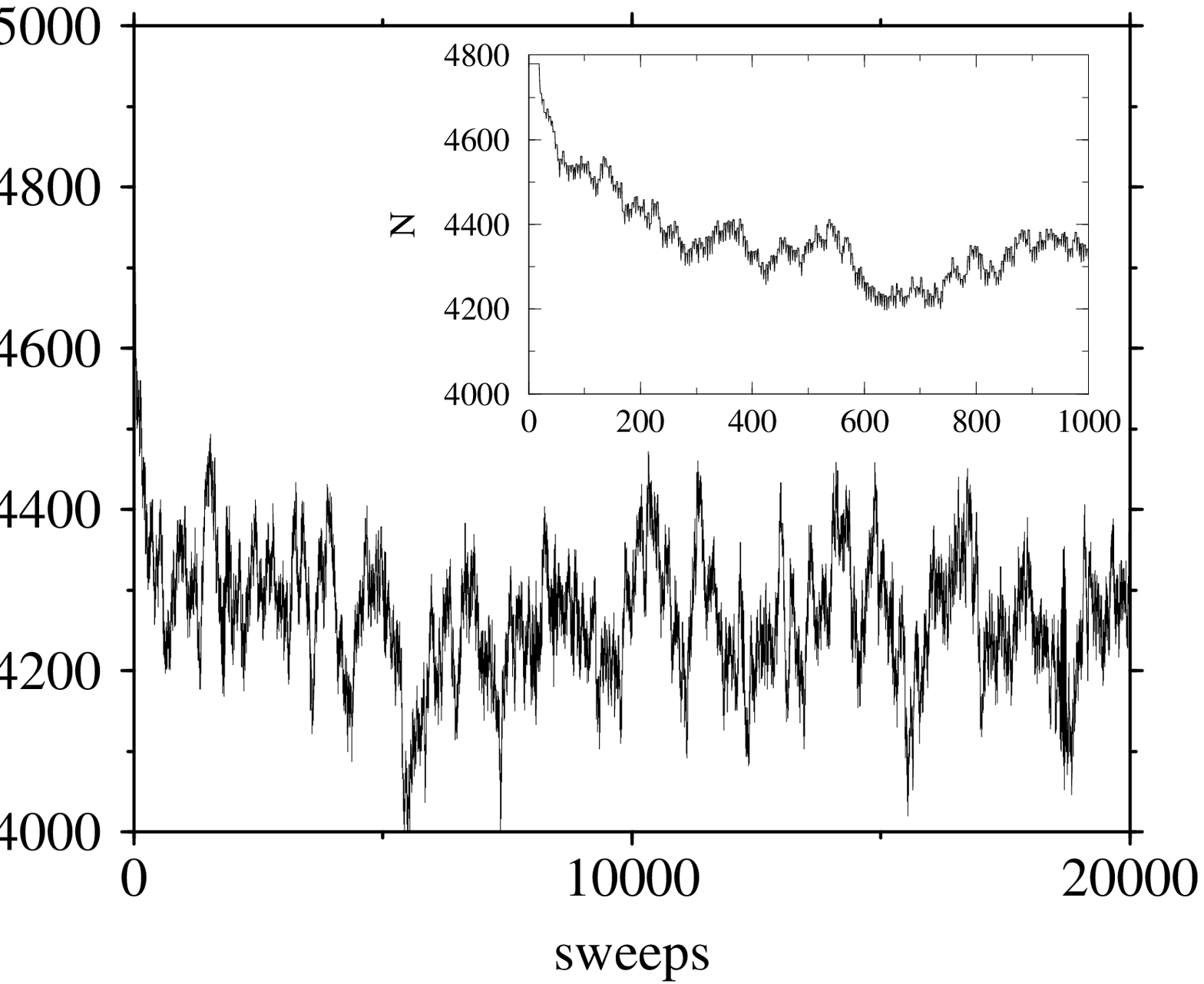} 
\hskip 0.2cm 
\epsfxsize 7.cm
\epsffile{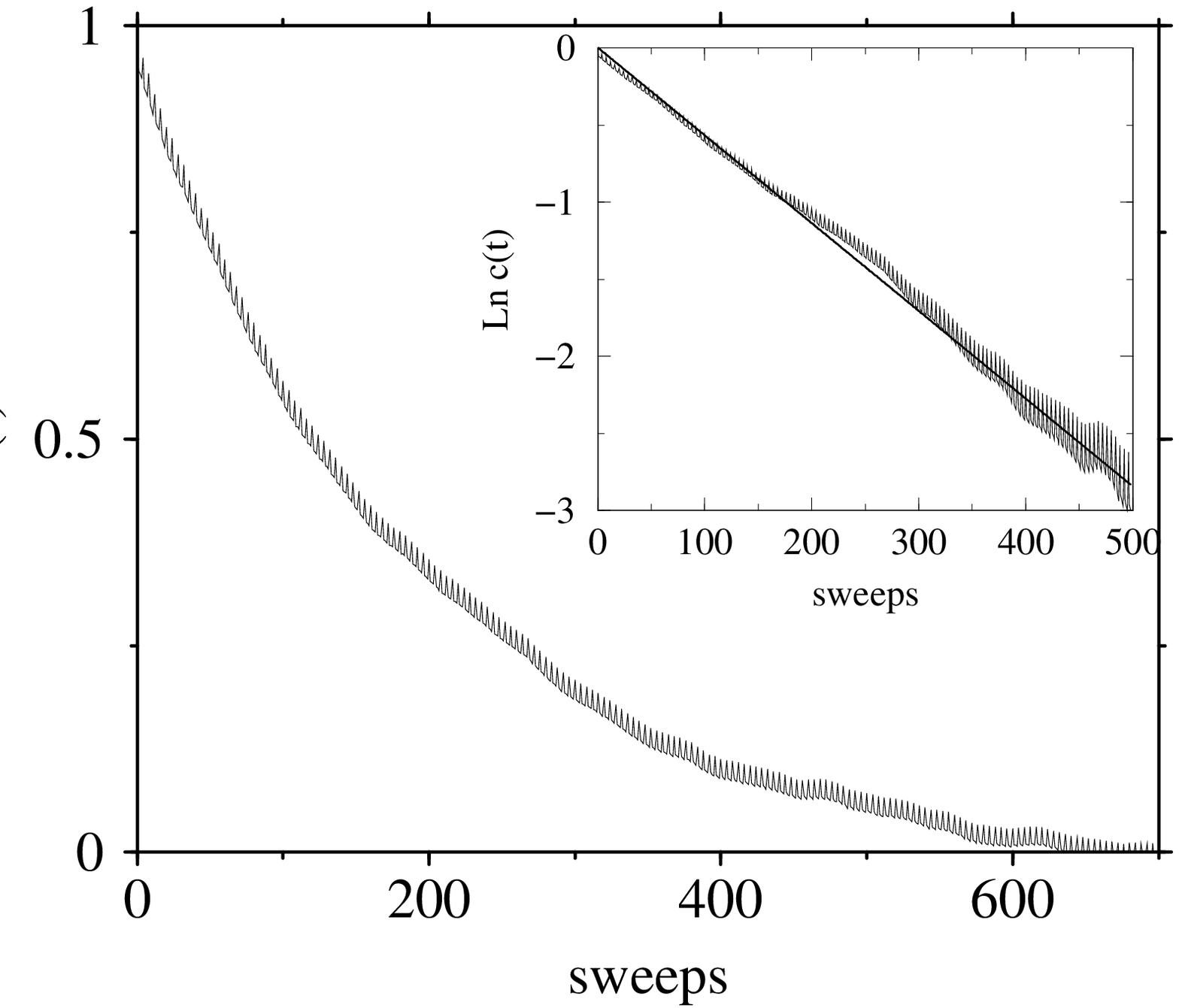}} 
\vskip 2cm
\caption{
U. Brodatzki, K. Mecke, Morphological Model for Colloidal Suspensions}  
\label{figtime_corr}
\end{figure}

In usual Metropolis Monte-Carlo Simulations the initial state converges
with a characteristic time scale $\tau$ 
towards a configuration sampled from the equilibrium
distribution. Whether or not such an equilibrium configuration is 
reached cannot be decided a priori and test runs have to be
performed, in particular, for this model where the evaluation of the
Boltzmann weights are difficult and extremely CPU-time expensive. In
Figure \ref{figtime_corr} we show a 
typical time  
series $N(t)$ of the number density of disks.  At each  sweep every
disk were tried to  move  exactly once. 
During the first 500
sweeps the systems relaxes from  the initial configuration of 4800
disks to the thermal equilibrium of $N=4272$ disks on average. The
relaxation time is found to be $\tau=180$ sweeps. 
At larger densities $N$ the relaxation time increases considerably, so
that equilibrium configurations could not be reached for densities
$x=\rho \pi R^2 > 3$. Therefore, it 
was not possible to verify the existence of a second fluid-fluid phase
transition at larger densities and a triple line predicted by a 
mean-field approximation of 
this model \cite{meckemorph}.

First, we use the algorithm to compute the average values of the Minkowski measures
$M_\nu(\rho)$ 
for $\beta=0$ as function of the density $\rho$ of the particles.  
 In Figure \ref{figphasediagram}(a) the mean
values  of  the covered area $M_0$ (full circles), the 
length  $M_1$ of the liquid boundary lines (triangles) and the Euler characteristic
$M_2$ (stars) are shown for three different ratios, namely  $\delta=0$ (dashed
lines),  $\delta=0.4$ (solid lines), and  $\delta=1.0$ (dot-dashed
lines). Whereas for  the pure hard-disk system ($\delta=1.0$) and for 
completely penetrable disks ($\delta=0.0$) a difference between the 
numerical results and theoretical values could not be found, one can
observe deviations from the analytic expression (\ref{minkhardcore1}) for large
densities $\rho\pi R^2 >2$ and $\delta=0.4$.  For small densities
$\rho \pi R^2$, the 
configuration ${\cal A}^N$ consists of isolated 
disks with negligible overlap of the wetting layers; therefore, each
measure starts out linearly in the number of particles. As the overlap
increases, the volume $M_0$ of the wetting layer saturates  and 
the  total boundary length $M_1$ and $M_2$ of the coverage decreases.
$M_2$ changes even its sign, since the curvature of the singular
corners (vertices) 
at intersections of the wetting layers is negative and starts to
dominate the positive 
contribution from the spherical parts of the boundary as the overlap
increases (see Figure \ref{decorated}). Negative values of the Euler
characteristic $M_2$ 
are typical for a highly connected structure with
many holes, i.e., undisturbed  cavities of binary fluid.

\begin{figure}
\centerline{\Huge Figure 3}
\vskip 4cm
\hbox{\epsfxsize 6.5cm
\epsffile{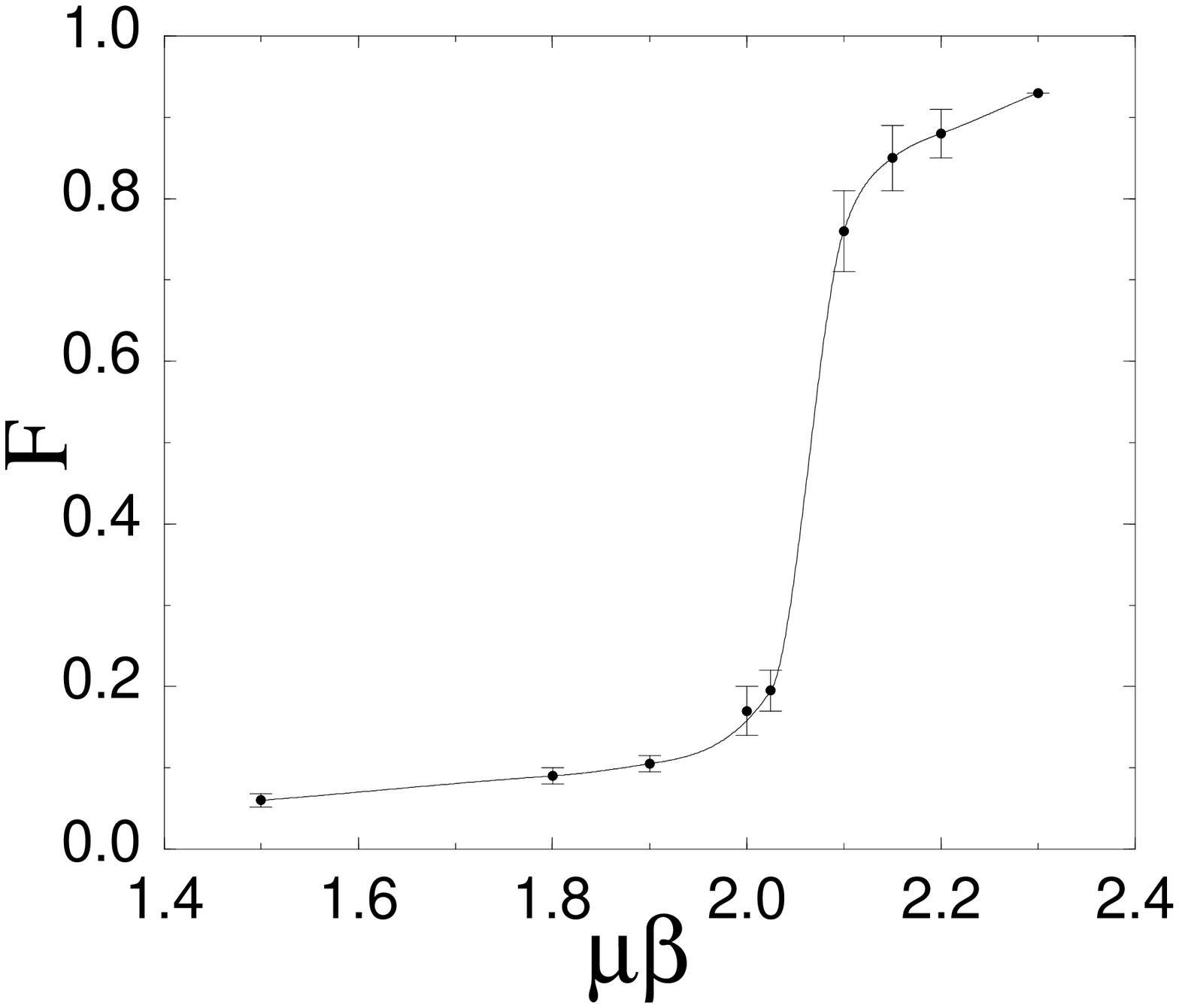}
\hskip -0.5cm 
\epsfxsize 6.5cm
\epsffile{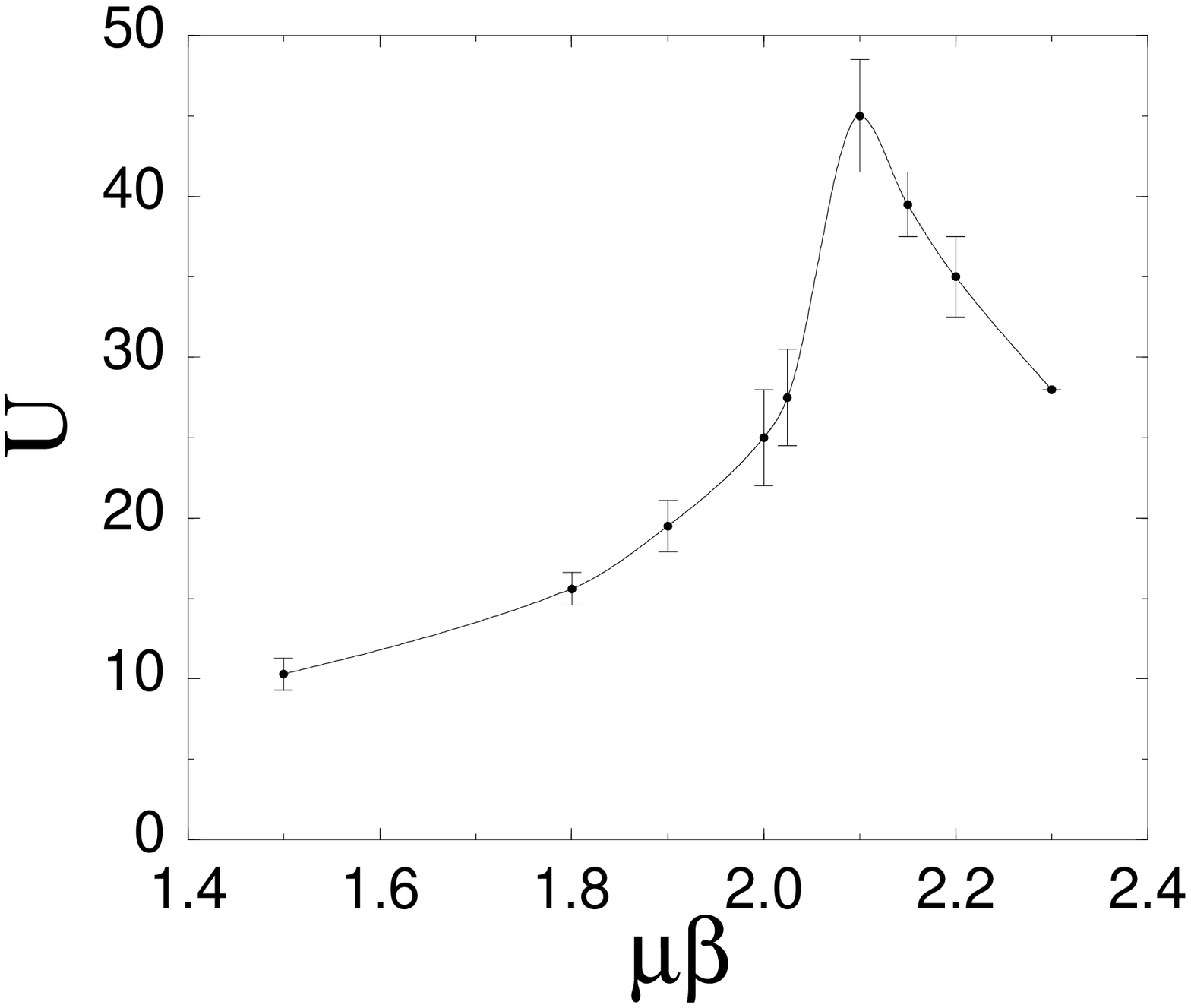}
\hskip -0.2cm 
\epsfxsize 6.5cm
\epsffile{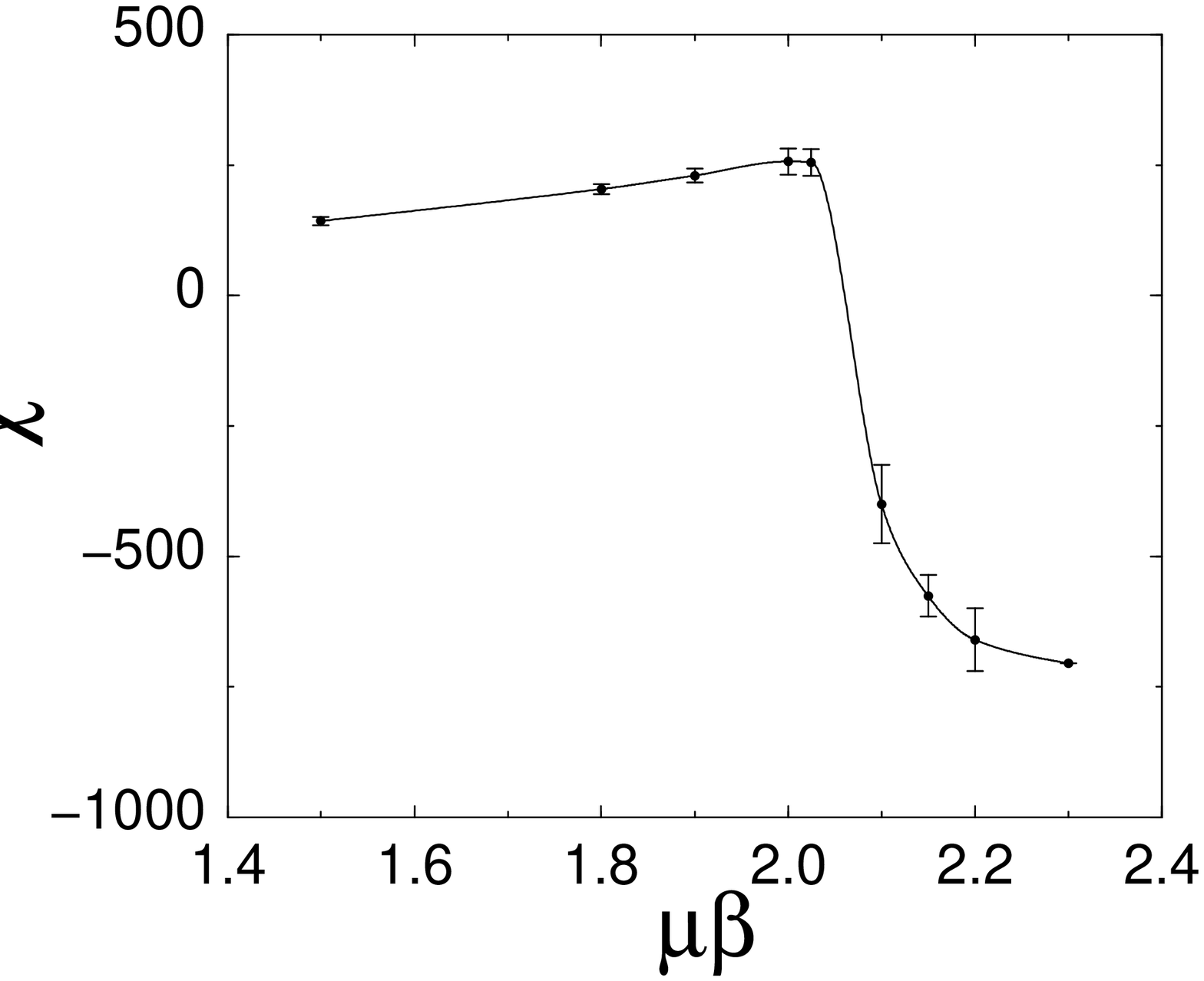}}
\vskip 2cm
\caption{
U. Brodatzki, K. Mecke, Morphological Model for Colloidal Suspensions}  
\label{figisotherm0.35}
\end{figure}

\begin{figure}
\centerline{\Huge Figure 4}
\vskip 4cm
\hbox{\epsfxsize 8.cm
\epsffile{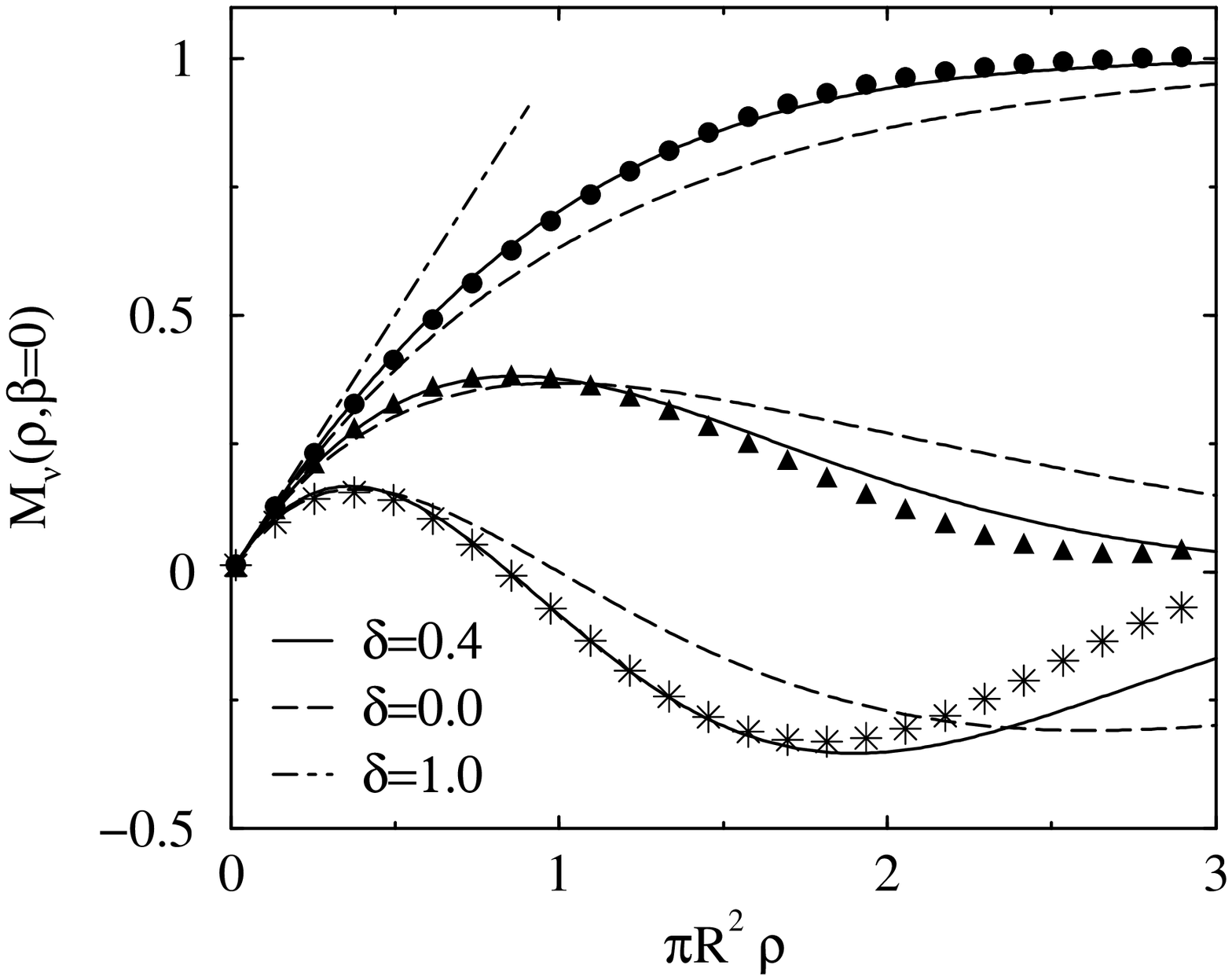}
\hskip 0.2cm   
\epsfxsize 8.cm
\epsffile{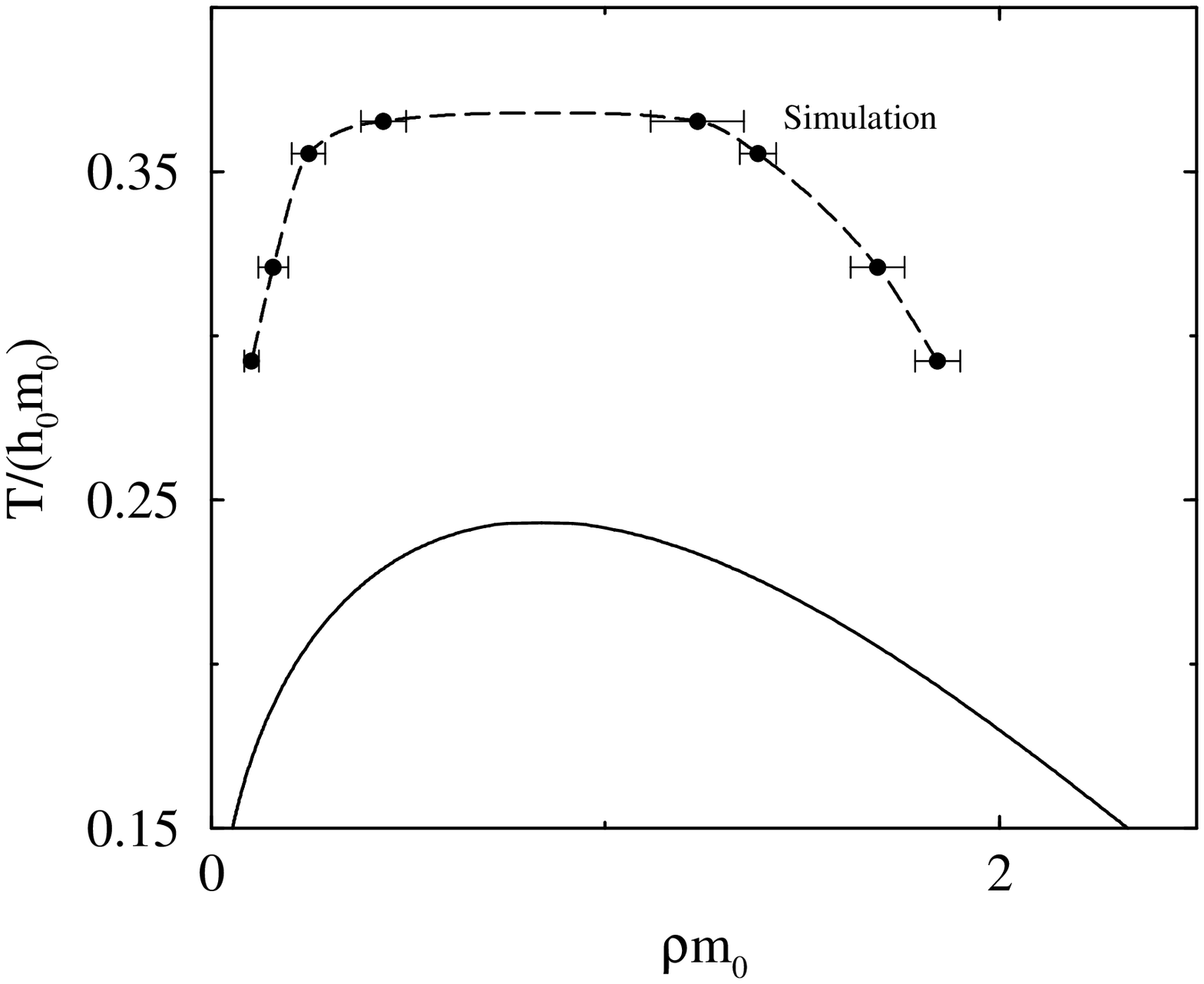}}
\vskip 2cm
\caption{
U. Brodatzki, K. Mecke, Morphological Model for Colloidal Suspensions}   
\label{figphasediagram}
\end{figure}

In Figure \ref{figisotherm0.35}  the morphological measures of the
wetting layer are shown  as function 
of the chemical potential $\mu$ at finite temperature $\beta h_0 \pi
R^2 = 1/0.355$.  
The discontinuities in the measures indicate clearly a phase transition from a
low density fluid phase (small volume $M_0$ of the wetting layer) to a
high density phase (large $M_0$) where the colloidal particles move
freely in a large region  formed by the union of the
wetting layers.  In 
Figure \ref{decorated} a configurational shnapshot close to the critical
point at $\tilde{T}_c=1/ (\beta \pi R^2 h_0) = 0.368$ and $\rho_c \pi
R^2 = 0.82$ is   shown, which illustrates both type of fluid phases. Note,
that the typical size of the connected wetting layer (white region) is much larger
than the distance  
between colloidal particles.

The coexistence of the low-density fluid phase of
isolated colloidal 
particles surrounded by a spherical wetting layer and the high-density
fluid phase 
where the colloidal particles move freely in a fluid  wetting phase
is shown  in Figure \ref{figphasediagram}(b).  
For the existence of the phase separation the inherent multi-particle
interactions of the colloidal particles are crucial. For $\delta =0$,
for instance, a pair-potential approximation of the Hamiltonian
(\ref{thermo1}) would not be bounded from below yielding unstable
thermodynamic configurations. And for $\delta << 1$ only the multi-particle
character of the Hamiltonian guarantees that for large densities
$\rho$ the 
interaction energy is limited so that the colloidal particles can move
freely within the fluid film.

Although the theoretical expression (\ref{minkhardcore1}) for the Minkowski
measures works well for densities $x<2$, the phase coexistence calculated
by perturbation theory is not in satisfying agreement with the
simulation results. Since the main goal of this approach is an
effective theory expressing the thermodynamic properties of the
colloidal systems in terms of the thickness $L(T,c)$ of the wetting
layer, one needs to improve the  perturbation theory before the
analytic results (\ref{minkhardcore1}) can be applied  to more 
realistic situations such as 
the experiments of silica beads in lutidine-water mixture  
\cite{beysens,gallagher,jayalakshmi}.

  The evaluation of the Boltzmann factor is very
CPU-time consuming which makes the 
simulation difficult, in particular, for large densities. An improved
algorithm could  verify the existence of a  second fluid-fluid
phase transition at larger densities 
which is indicated by  perturbation theory  due to multiple 
particle interactions of the Euler characteristic in the Hamiltonian.

\newpage

\noindent
\centerline{\Huge Figure captions}

\bigskip 
\bigskip 
\bigskip 

{ \Large Fig.1:} 
An ensemble of
hard colloidal particles (black disks) surrounded by a fluid wetting 
layer (white ring). The interaction between the colloids is 
determined 
by  the free energy of the  fluid film
(white region) 
which may cause  a fluid-fluid phase separation 
of the hard particles. 
Thus, the spatial structure of the phases, i.e., the morphology of the
white regions determines the configurational energy which determines
itself the  spatial structure due to the Boltzmann factor in the
partition function of a canonical ensemble. A 
main feature of this model is the occurrence of different
length scales: the clusters of the particles, i.e., the connected
white regions are much larger than the 'microscopic' radius of the
disks and the typical nearest-neighbor distance within a cluster.

\bigskip 
\bigskip 
\bigskip 

{ \Large Fig.2:} 
Times series $N(t)$ and corresponding correlation function
$c(t)$ where the first 3000 sweeps were omitted. The temperature
$\tilde{T}=0.36$ is slightly below the critical point at
$\tilde{T}_c=0.368$ for $h_0=1$, $h_1=h_2=0$, where $\tilde{T} =
k_BT/(\pi R^2 h_0)$ denotes the reduced temperature.

\bigskip 
\bigskip 
\bigskip 

{ \Large Fig.3:} 
Isotherms at $\tilde{T}=0.355$ ($\delta=0.4$, $h_0=1$,
$h_1=h_2=0$) of the density 
$\rho$, the covered area $F=M_0$, the 
length  $U=2\pi M_1$ of the liquid boundary lines and the Euler characteristic
$\chi=\pi M_2$ measuring the topology of the configurations. One can clearly
observe the phase
transition at $\beta\mu=2.1$. The functional form of the morphological
measures are typical: $\chi$ exhibits a discontinuous jump from positive to
negative values, $U$ has its maximum at the transition, whereas $F$ is
a monotonous increasing function.

\bigskip 
\bigskip 
\bigskip 

{ \Large Fig.4:}  
(a) Isotherms at $\beta=0.0$  and $\delta=0.4$ of  the covered
area $M_0$ (full circles), the  
length  $M_1$ (triangles) of the liquid boundary lines  and the Euler characteristic
$M_2$ (stars). The analytic result for $\delta=0.4$ (solid lines) given by Eq. (\ref{minkhardcore1}) works quite well for
densities $x<2$. For $\delta=0.0$ (fully penetrable disks, dashed
lines) and for $\delta=1.0$ (hard disks, dot-dashed line) no
differences between numerical determined values and theoretical
predictions could be found. (b) Phase diagram with $h_1=h_2=0$, $h_0=1$, and
$\delta=0.4$; $m_0=\pi R^2$.  The
 dashed line indicates the location of the coexisting densities. The 
perturbation result (solid line) based on the isotherms (a)
overestimates the stability of the homogeneous fluid 
phase yielding a lower critical temperature.

\newpage

\vfill\eject 

\newpage

\newpage

\newpage

\end{document}